# DBR: A Simple, Fast and Efficient Dynamic Network Reconfiguration Mechanism Based on Deadlock Recovery Scheme

Majed ValadBeigi, Farshad Safaei and Bahareh Pourshirazi

Department of Electrical and Computer Engineering, Shahid Beheshti University G.C, Evin 1983963113, Tehran, IRAN
M_valadbeigi@sbu.ac.ir, f_safaei@sbu.ac.ir and
b.pourshirazy@mail.sbu.ac.ir

**ABSTRACT**

*Dynamic network reconfiguration is described as the process of replacing one routing function with another while the network keeps running. The main challenge is avoiding deadlock anomalies while keeping limitations on message injection and forwarding minimal. Current approaches, whose complexity is so high that their practical applicability is limited, either require the existence of extra network resources like virtual channels, or they affect the performance of the network during the reconfiguration process. In this paper we present a simple, fast and efficient mechanism for dynamic network reconfiguration which is based on regressive deadlock recoveries instead of avoiding deadlocks. The mechanism which is referred to as DBR guarantees a deadlock-free reconfiguration based on wormhole switching (WS) and it does not require additional resources. In this approach, the need for a reliable message transmission has led to a modified WS mechanism which includes additional flits or control signals. DBR allows cycles to be formed and in such conditions when a deadlock occurs, the messages suffer from time-out. Then, this method releases the buffers and channels from the current node and thus the source retransmits the message after a random time gap. Evaluating results reveal that the mechanism shows substantial performance improvements over the other methods and it works efficiently in different topologies with various routing algorithms.*

## 1. INTRODUCTION

Computers get faster, but the demand for more computing resources seems to grow at an even faster rate, and depending on the applications domain, this demand can be satisfied by either, massively parallel computers, or cluster of computers. The dependency on high performance interconnect networks such as Myrinet [1], Infiniband [2], Gigabit Ethernet [3, 4], and Quadrics [5] is common for both approaches.

Interconnection networks are applied as the communication infrastructure of parallel processing systems which enable the diverse processing, memory, storage, and I/O components of system to communicate.

They are found in high-end servers [6] in the form of system area networks [7] as well as in multicore processors [8] as networks-on-chip (NoCs) [9] at the other end of the spectrum. The network's role is critically relying on determining the system performance and dependability as the interaction and cooperation of other system components; ultimately it can be said that the role of the network depends on its ability to establish communication paths between the above-mentioned components [10].

Various switching mechanisms have been described in the literature for interconnection networks including *packet switching (PS)*, *virtual cut-through (VCT)* and *wormhole switching (WS)* [12].

WS [11] (also referred as wormhole-routing) has become the most widely used switching mechanism for multicomputers and distributed shared-memory multiprocessors, and it is also being used for networks of workstations [1]. Besides, a message in WS is fragmented into elementary units, called *flits*, for transmission and flow control [12].

In PS and VCT, messages are completely buffered at a node. As a result, the messages consume network bandwidth proportional to the network load. On the other hand, wormhole-switched messages may block the occupying buffers and channels across multiple routers, precluding access to the network bandwidth by other messages [12].

The need for a reliable message transmission has also led to a modified WS mechanism which includes additional flits or control signals (e.g., acknowledgments or padding flits). This particular technique was proposed as *compressionless routing* by its developers [13]. By increasing the probability of failure and reliability concerns for interconnection networks, fault-tolerance has quickly become an indispensable part of such systems. Thus, it is necessary to provide an efficient fault-tolerant mechanism to keep the system running despite the presence of faults is necessary.

Fault-tolerance is defined as the ability of a system to pursue an operation, even in the presence of faults [14]. *Reliability*, *availability* and *dependability* are the three most applicable terms of fault-tolerance [15]. However, due to the large application area, interconnection networks are found in systems with high requirements for reliability and continued operation.

The use of fault-tolerance mechanisms will assure that in case of a component failure the system keeps working, although in a degraded mode it should wait until the failed component is repaired. Basically, there are three ways to cope with faults in the interconnection networks: *component redundancy*, *fault-tolerant routing algorithms* and *reconfiguration techniques* [16]. Using the component redundancy has been the easiest and a costly way. In this method, while a failed component is detected in the system, it is easily replaced by its redundant copy.

Fault-tolerant routing algorithms aim at preventing messages from traversing faulty components by providing some kinds of routing path redundancy. To reach this end, messages must be able to be routed through alternative paths to circumvent or avoid faulty regions over the network. Fault-tolerant routing schemes should be designed to tolerate a certain number of faults while still guaranteeing deadlock freedom in the network. However, to fulfill the requirements, fault-tolerant routing strategies often need to use additional network resources such as virtual channels or additional hardware at switches or routers.

By applying reconfiguration [17], any number of faults can be tolerated while the network remains physically connected. Once a fault is detected, the configuration consists of a fault is identified and the new topology will be discovered. Thus, a new routing scheme is computed and the required components in the network are updated. The main disadvantage of reconfiguration is the high delayed messages that may occur during the reconfiguration process.

*Reconfiguration* techniques can be either *static* or *dynamic*. Static reconfiguration techniques require the network traffic to be completely stopped the traffic in the network before changing any routing table, so the network is emptied. The routing algorithm used after the reconfiguration process is different. It implies that all the paths for each source-destination pair need to be computed. Owing to the network down-time, i.e. halting message injection that may cause strong performance degradation during the reconfiguration process, static reconfiguration largely impacts on the message latency. This issue prevents static reconfiguration techniques from being used in systems with high performance requirements.

Unlike the static reconfiguration, in a dynamic reconfiguration the transition from one routing function to another is performed while the functional parts of the network are fully operational, i.e., we have no network down-time and no halting message injection. This typically leads, when compared with static reconfiguration, to a reduction in the number of messages that miss their quality of service deadline. The problem in this approach resides in the fact that, in general, two different and individually deadlock-free routing functions may be prone to deadlock if they coexist in the network. It means that, in a dynamic reconfiguration, there will be a transition phase between the old and new routing functions where *reconfiguration-induced deadlocks* may occur. Another drawback of using dynamic reconfiguration is that it usually requires extra resources.

In this article, we introduce a simple, fast and efficient method for dynamic network reconfiguration which is based on regressive deadlock recoveries instead of avoiding deadlocks. DBR guarantees a deadlock-free reconfiguration based on WS and it does not require additional resources. In this approach, the need for a reliable message transmission has led to a modified WS which includes additional flits or control signals. This can be achieved by padding messages [13]. Further, a message can not leave the source node until the header flit reaches its destination. Moreover, deadlock recovery is basically achieved through using time-out mechanism [18]. DBR allows cycles to be formed and in such conditions when a deadlock occurs the messages suffer from time-out. This releases buffers and channels at the current node containing the header that goes back to the source node by reversing back along the path reserved by the header message and thus the routing table is updated. When a message experiences a transmission failure, due to a time-out at an intermediate node, the source retransmits the message after a random time gap.

The rest of the paper is organized as follows. Related work is presented in section 2. In Section 3, we present DBR method and describe implementation details. In Section 4, DBR is evaluated. Finally, in Section 5, conclusions are provided.

## 2. RELATED WORK

Faults in a network appear in several different types, such as hardware faults, software bugs, or malicious sniffing or removal of packets. The first step in dealing with errors is to understand the nature of component failures and then to develop simple models that allow us to reason about the failure and the methods for handling it. Classification of faults by nature is either *random* or *systematic* faults. Random faults are usually hardware faults affecting the system components occurring with a certain probability, while systematic faults such as software failures are the faults which are not random, whether a component has it or not [12]. We assume that such permanent failures are detected and contained of a node or a link boundary. Thus, faults are assumed to be fail-stop [19], meaning that we do not consider Byzantine (i.e., malicious) faults [12]. In the contexts of fault-tolerant routing, these are common assumptions [12, 19-20].

Faults also can be classified by their duration as *transient* and *permanent* faults [12]. Transient faults stay in the system for only a short duration, while permanent faults remain in the system until it is repaired. Permanent faults may be either *dynamic* or *static*. In a dynamic fault model, while a new fault is found, actions are performed in order to appropriately handle the faulty component which allows the system to reconfigure at the hardware level, and preserves the original network topology.

In some situations the defined promises of the routing algorithm and/or network topology may break, affecting the network dependability. This may happen when the topology of the network changes, either involuntarily due to faulty components or voluntarily due to removal or addition of some components. This normally requires the network routing algorithm (routing function) to be *reconfigured* in order to re-establish the connectivity of the entire network [22].

Unlike static reconfiguration techniques, dynamic reconfiguration techniques [17] do not require the network traffic to come to a complete stop. However, some packets must be removed from the network and re-injected later, which could cause a strong degradation in performance during the reconfiguration time. In the last decade several dynamic reconfiguration mechanisms have been proposed. Next we describe some of them.

In [17], a *Partial Progressive Reconfiguration* (PPR) technique is proposed, allowing arbitrary networks to migrate between two instantiations of *up\*/down\** routing. The effect of load and network size on PPR performance is evaluated in [23].

Another approach is the *NetRec* scheme [24] which requires every switch to maintain information about the switches in some hops away. Yet another approach is the *Double Scheme* (DS) [25] that uses two sets of virtual channels in the network which act as two disjoint virtual network layers during the reconfiguration. A methodology for deriving new reconfiguration processes for any given pair of old and new routing function is given in [22]. An orthogonal approach which may serve on top of all above techniques is explained in [26], where, for up\*/down\* routing only some parts of the network need to be reconfigured for up/down routing. Solid theoretical supports on the issue that dynamic reconfiguration design methodologies and techniques are proved to be deadlock-free, can be found in [27].

Moreover, a mechanism was suggested in [22] which is referred to as *Simple Reconfiguration* (SR). In SR a token is issued to separate the messages routed with the old routing function from messages routed with the new routing function. Tokens advance through an output port in a switch once there are no more old messages passing through the output port (based on input and output dependencies generated from the old routing function). By performing this, there are no cycles in the network since there will be no old messages behind new ones.

The above mentioned mechanisms lack at least one of the identified goals in this paper. In particular, PPR only works with routing functions that adhere to the up\*/down\* scheme. NetRec [24] is specially tailored for re-routing messages around a faulty node. It fundamentally provides a protocol for generating a tree that connects all neighbor nodes of a fault, and drops packets to avert deadlocks in the reconfiguration period. DS is more flexible, in the sense that it can handle any topology and transition between any pair of deadlock-free routing functions. However, it requires the presence of two sets of virtual channels. The methodology in [22] requires complex computation in order to derive a safe reconfiguration process once the new routing function has been chosen. It consumes time and thus limits the applicability of the methodology. SR mechanism requires a token to be distributed over the entire network. Although it separates old and new traffic, it has two major drawbacks. The first one is that its implementation is not straightforward. The token distribution is based on the dependencies of the old routing function. The second, messages suffer from extra blocking since new messages must wait for the tokens to advance.

DBR can exhibit superior performance characteristics over the following goals. First, messages are not getting blocked for any reason. They are routed as soon as possible. Therefore, the message latency is minimized. Second, the nodes which are closed to failed links or nodes are updated in a fast manner. Third, the mechanism does not require additional resources at network components. Fourth, nodes react quickly in the presence of a reconfiguration process. The mechanism has been performed based on WS and works with any routing algorithm implemented in the form of routing tables at nodes.

It is worth mentioning that the authors in [21] suggested a protocol for dynamic network reconfiguration mechanism referred to as PDR in order to handle both deadlock and performance degradation. PDR provides an efficient approach for both deadlock detection and deadlock recovery. Further, although it has many advantages such as superior performance in higher message injection rates and simplicity, it needs additional resources such as virtual channels to pursue its operations. It is worthy to mention that in Section 4, we have compared our method with DS and SR. Also in [21] the method of PDR has been compared with the two recent mechanisms. We have avoided the comparison between PDR and DBR. The reason is the fact that both DS and SR mechanisms are based on deadlock avoidance, so the suggested technique of PDR is a mechanism based on deadlock recovery and a comparison between this mechanism and the proposed method in this paper, i.e. DBR, involves considering other parameters of performance and method comparisons of deadlock detection and recovery that is far from the main focus of this paper. Therefore, the comparison between these two mechanisms has been postponed to another separate paper.

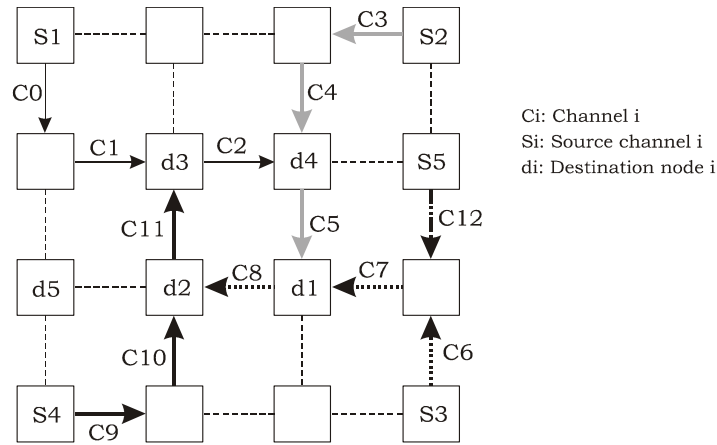

Figure 1. Cycle in a dynamic network reconfiguration. Messages labeled by $\hat{A}_{New}$ are sent from S1, S3 to d1, d3, respectively while messages labeled by $\hat{A}_{Old}$ are sent from S2, S4, S5 to d2, d4, d5, respectively.

## 3. DBR MECHANISM

We consider the use of two routing algorithms referred to as $\hat{A}_{Old}$ and $\hat{A}_{New}$. Routing information is distributed along the nodes by using routing tables. The mechanism is based on the fact that deadlock does not occur frequently, so recovery may be preferable to prevention. Indeed, the probability of deadlock is proportional to the traffic injection rate and inversely proportional to the availability of resources.

The DBR mechanism will allow cycles to be formed (Figure 1). In this situation, the mechanism is used to remove the deadlock by releasing the reserved path (routing table). The basic idea is to remove the deadlock by releasing the reserved path (routing table). In this situation, the fine-grained flow control and backpressure of wormhole-routing is used to communicate to nodes the routing tables, the routing status and the error condition. The nodes use the information to provide deadlock avoidance. In DBR, deadlock is avoided by keeping track of whether the message header has reached the destination or not. If it reached, no deadlock is possible; otherwise, it is blocked for a particular time and then the source tears down the partial message path (routing table) and tries again later. Thus, any deadlock message ($Â_{Old}$ or $Â_{New}$) will eventually have its path torn down.

To determine whether the message header has reached the destination, DBR takes the advantage of the properties of messages under wormhole routing. Wormhole routing provides feedback in the form of flow control which can be exploited to communicate acknowledgements. Messages have a fixed profile in channels due to the small amount of buffering in wormhole routers. So when the message is long enough, the sender can determine that the message header must have reached the destination if a sufficient number of flits have been injected into the network. Otherwise, the sender pads the message to ensure that the header reaches the destination before the last flit has been injected by the source. When the real data ends (at the rising edge of the pad signal), the receiver will be informed by the pad signals. Else, the reserved path ought to be released (at the falling edge of the pad signal) [13]. Figure 2 demonstrates the single message format in DBR. For the sake of explanation, DBR is described as a fully deployed mechanism following the entire reconfiguration process from the occurrence of the topological change (i.e., a failure) to the normal and final functioning of the network with the new routing algorithm. The following sections describe each step.

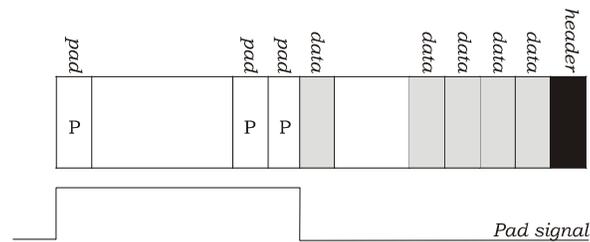

Figure 2. The message format for DBR mechanism [13]. The pad signal is used to differentiate pad and data flits.

### 3.1. Status Information Distribution

In some situations the defined promises of the routing algorithm and/or network topology may break and naturally affect the network dependability. This may happen when topology of the network changes, either involuntarily due to faulty components or voluntarily due to removal or addition of components. This normally requires the network routing algorithm to be reconfigured in order to network connectivity to be re-established. DBR is applied whenever a new routing algorithm is needed for the network. In some conditions the change in the topology does not necessarily lead to a change in the routing algorithm. This is the case of adding/removing nodes to/from the system. As the routing algorithm remains unchanged, there is no probability of deadlock and no demanding for a global reconfiguration process. On the other hand, a new node or link might be added, changing the topology. In that case, a change of the routing algorithm could lead to achieve a higher performance and thus activating a reconfiguration process. The routing algorithm might also need to be changed even if no topological change occurs. In cases which the topology makes no changes in, with changing the routing algorithm higher performance might be achieved. However, the most considerable topological change is when a node or link fails (or a group of them). Once some parts of the network are being disconnected, altering the routing algorithm is required. To alleviate this problem, we will introduce an efficient mechanism in the situation that a failure occurs.

In all the above mentioned cases, DBR reconfiguration mechanism is activated. To do so, a selected node runs a component to be responsible for detecting any topological change. To achieve this, the component sends control signals periodically to all nodes. Nodes respond to the current status of their links and neighbor nodes. On detecting the topological change, new routing tables need to be computed. At the moment that the nodes are notified, they use the alternate paths to bypass the failed node. However, employing the alternate paths is an additional mechanism which has nothing to do with the reconfiguration process. Therefore, we only focus on the reconfiguration process (i.e., updating tables).

Once the paths are computed, they must be distributed by sending all the new routing tables to all nodes through the network. Figure 3 shows an example of the sequence used to updating the nodes. Once a node finds out the new paths, it updates its routing table removing the old one.

Additionally, in order to reduce the overhead control traffic, only the differences between the old and the new routing tables might be sent to every node. Depending on the similarity of the routing algorithms the percentage of control traffic reduction may be significant. Indeed, in the evaluation we will see that this improvement will affect the effectiveness of the mechanism.

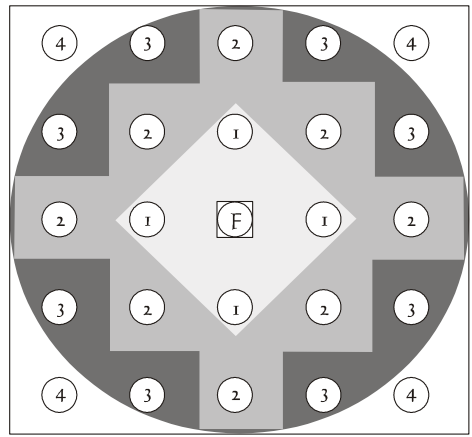

Figure 3. Sequence updating of routing table while the failure occurs at node F. The numbers indicate both the sequence and the distance to the failed component.

### 3.2. Deadlock Detection and Recovery

In order to keep track of paths that potentially can be released to break deadlocks, DBR detects deadlock by using a time-dependent selection function similar to those suggested in [18]. Besides, DBR exploits the tight coupling of wormhole routers for flow control to perform deadlock recovery. The detected deadlock is recovered by releasing the path.

To send a message, the sender first resets two initial parameters $F$ and $C$ which are referred to as flit counter and blocking counter, respectively. The former indicates how many flits of the current message have been injected while the latter states how long the message header has been blocked at a node. We also introduce the minimum flit injection for delivery guarantee parameter which is depicted by $F_{path}$. $F_{path}$ is defined as depth of channel buffer (flits/channel) × distance to destination in hops.

The parameter $F$ is incremented in each cycle when a new flit is injected and $C$ is added up in each cycle when a flit cannot be injected. If the message in flit ($F$) is shorter than the distance in flits ($F_{path}$), it is padded to make the size equal to the distance in order to reserve the path. If the sender is unable to send out the header, $C$ is incremented. Incrementing the parameter $C$ on every cycle by the router is continued until the router succeeds to send out the header or the value of $C$ reaches the time-out interval (denoted by $T$) for that message. When $C > T$, the router changes its status to "deadlocked" meaning that a cycle has been formed.

If there were cyclic dependencies (during transition from $Â_{Old}$ to $Â_{New}$) among the channels in the network, there would be no path to escape from cycles and the sender cannot inject a new flit for a period longer than $T$. It shows a deadlock situation, so that the sender launches a release signal to release the path. The $F$ and $C$ are also reset and the same message will be re-injected later.

## 4. EVALUATION

In this Section we evaluate the proposed reconfiguration mechanism. To do so, we first present the evaluation methodology and traffic patterns. Then, we briefly describe the reconfiguration mechanisms used for comparison purposes. Finally, results and analysis are presented.

After the text edit has been completed, the paper is ready for the template. Duplicate the template file by using the Save As command, and use the naming convention prescribed by your conference for the name of your paper. In this newly created file, highlight all of the contents and import your prepared text file. You are now ready to style your paper; use the scroll down window on the left of the MS Word Formatting toolbar.

### 4.1. Evaluation Methodology

We have developed a detailed simulator that allows modeling the network at the cycle level. An event-driven simulator as Xmulator [28] was used for evaluating the performance of the proposed methodology. The Orion power library [30] is integrated to our simulator to calculate the power consumption of the networks.

In order to determine the fault-tolerance of our methodology's variations, we have performed dynamic reconfiguration analysis for an 8×8 torus topology. WS is considered as the flow control mechanism. The simulations have been performed using a base message size of 16 flits and also the width of 128 bits for each flit. Moreover, each physical link is split into two virtual channels (VC). We calculated the power consumption of the links of each router in 70 nm technology library. In this technology, the clock frequency is set to 250 MHz, and the length of links between two adjacent routers is set to 1 mm for the torus topology. We also have performed a large number of simulations in order to make the evaluation results independent of the relative positions of faults.

### 4.2. Traffic Pattern

We consider two different traffic patterns when evaluating the network behavior: *synthetic patterns* and *traces* [14]. Synthetic patterns are widely used because they allow evaluating the network in the most generic way. When we use them, every node has the same traffic injection rate. We evaluate the complete range of traffic injection rate, from low levels up to the saturation point. The used synthetic traffic patterns are uniform and Hotspot [14].

International Journal of VLSI design & Communication Systems (VLSICS) Vol.3, No.5, October 2012

- For uniform traffic, each source node sends messages to all the destinations with the same probability.
- For hotspot, 10% of the sources (selected randomly) inject traffic to the same destination (selected randomly), the rest of end nodes inject traffic to random destinations. This traffic pattern allows to model the situation when one or more end nodes are frequently accessed by the remaining end nodes (a disk server, for instance).

On the other hand, traces are based on capturing the traffic when running real applications. Traces contain the source, destination, injection time and the size of each sent message. They allow obtaining results in more realistic scenarios and let us compare them with the results obtained when using synthetic patterns. In this paper, some results obtained with this type of traffic pattern are shown.

The used traces were extracted under the execution of the *FFT, LU, BARNES, RADIX, WATER-Nsquared and WATER-Spatial* applications from SPLASH-2 [29] suite in shared-memory multiprocessors. These types of applications are widely used when simulating multiprocessor systems on engineering and scientific computations.

### 4.3. Evaluation Mechanism

DBR mechanism is evaluated when sending all the routing tables. We compare DBR with SR and DS mechanisms.

In all the reconfiguration mechanisms, once a topology change is detected, the new routing tables along with a control message are sent to all the nodes through the control virtual channel. In the case of DS, during the distribution of paths one virtual channel is drained. Once drained, control messages are sent to restore normal operation (the reconfiguration has finished). In the case of SR, the tokens that separate the old and the new traffic at the same time are sent to the nodes. More details of DS and SR can be found in [10].

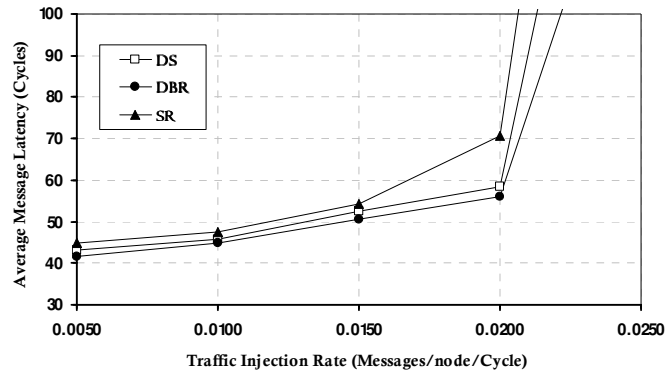

Figure 4. Average message latency vs. traffic generation for an 8×8 torus network under the uniform traffic pattern.

### 4.4. Results and Analysis

We evaluate all the reconfiguration mechanisms for a random node failure in an 8×8 torus network. In this case, the old and the new routing algorithms are the up*/down*.

Figure 4 shows the average message latency for the 8×8 torus for each reconfiguration mechanism and for different injection rates. The figure reveals that the latency of SR scheme increases significantly.

The reason is that SR experiences a higher latency due to the blocking introduced by the tokens. This effect is rapidly extended and thus messages experience higher latencies that this phenomenon carries on even though the reconfiguration process has finished. Further, DBR illustrates better results compared to DS and SR. Figure 4 confirms that DBR shows the superior performance to three different injection rates due to the structure of the approach as explained earlier.

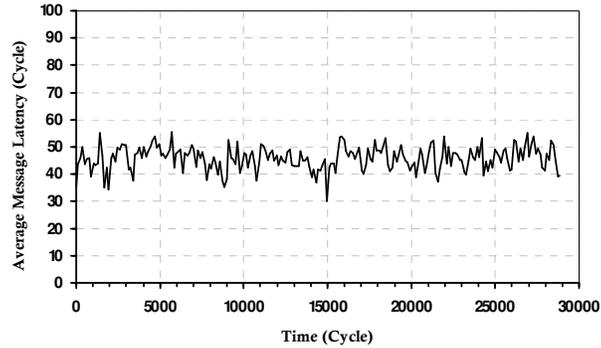

(a) DBR

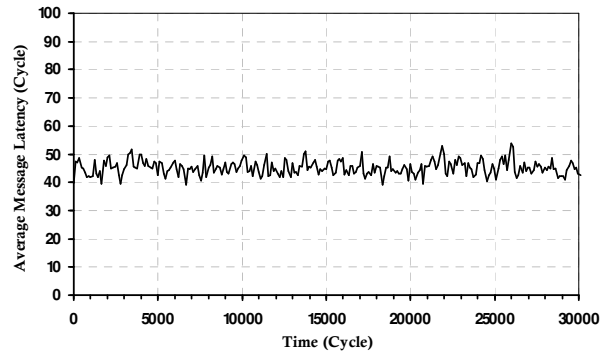

(b) DS

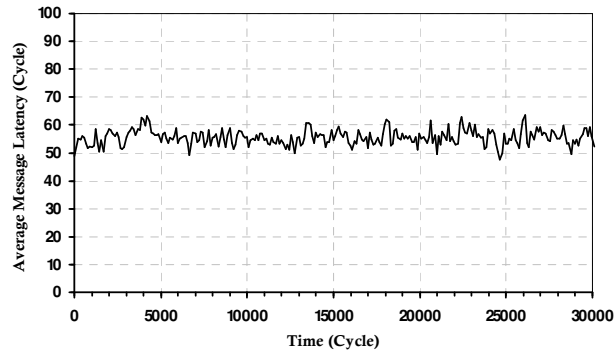

(c) SR

Figure 5. Average message latency in an 8×8 torus network during the reconfiguration process.

Moreover, each mechanism requires the same reconfiguration time regardless of the traffic rate. This is because in all the cases the distribution of routing tables is the process that takes most of the time (tables are sent sequentially) and the amount of information to distribute is the same despite the traffic injection rate. Also, it is apparent that control messages use the reserved control virtual channel and they have higher priority than data messages.

Figure 5 shows the average message latency plotted against the generation time for the different reconfiguration schemes under the uniform traffic. As can be seen in Figure 5(a), there are two interval times (from 1,000 to 2,000 and 15,000 to 16,000 cycles) on which the curve sharply increases. The reason behind this behavior is rooted in the nature of spending extra time for releasing the path when the deadlock is detected and re-injecting the message. Yet, DBR shows even better results in average message latency overall.

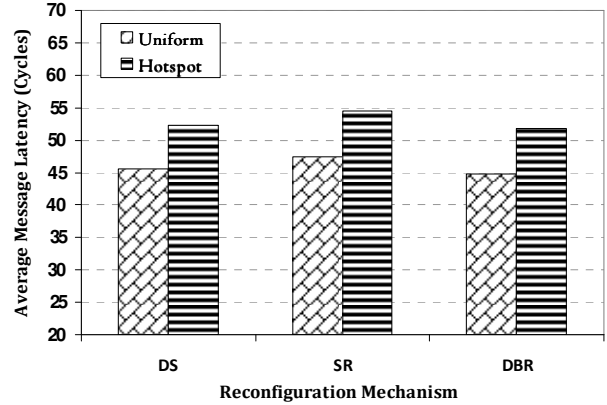

Figure 6. Average message latency in an 8×8 torus network under different synthetic traffic patterns.

We have also evaluated the mechanisms with a variety of traffic patterns. Figure 6 shows the average message latency for an 8×8 Torus network for different synthetic traffic patterns.

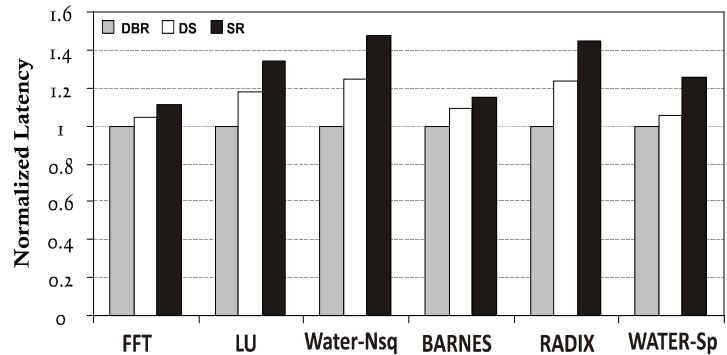

Figure 7. Normalized latency in DBR, DS, and SR for SPLASH-2 programs.

Figure 7 compares the latency by DBR, SR, and DS across six SPLASH-2 traces, on a 7×7 torus. The results are normalized to the results given by DBR. On average, DBR outperforms DS by 14% and SR by 29% when considering the message latency. For the programs such as *WATER-Nsquard* where the traffic is distributed rather evenly across the nodes, DBR provides a better improvement compared to DS over the SR.

Figure 8 compares the Power obtained by DBR, SR, and DS across six SPLASH-2 traces, on a 7×7 torus. The results are normalized to the results given by DBR. On average, PDR outperforms the DS by 4% and SR by 12% when considering the Power Consumption.

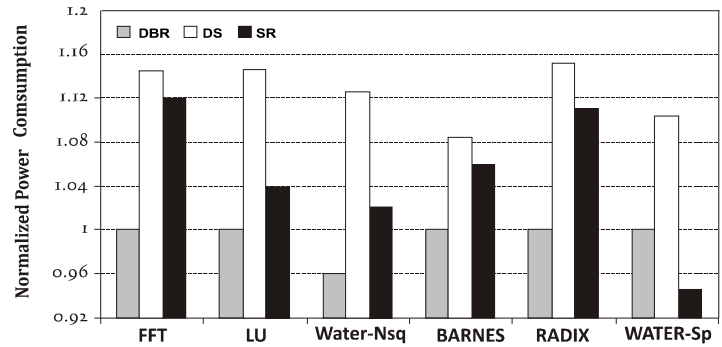

Figure 8. Normalized power consumption in DBR, DS, and SR for SPLASH-2 programs.

To sum up, Table 1 illustrates the diverse conclusions we can extract from the performed evaluations. As can be seen DBR mechanism achieves the best results in the evaluation parameters mentioned in the table except the complexity which DS shows the better result.

TABLE 1. Summary of the characteristics of the different dynamic reconfiguration mechanisms

| Reconfiguration Mechanism | Resources Needed | Complexity | Messages Dropped | Traffic Perturbance | Power Consumption |
|---|---|---|---|---|---|
| DS | 2 VCs | Low | High | No | Medium |
| SR | tokens | High | High | Yes | High |
| DBR | – | Medium | Low | No | Low |

As an overall conclusion it can be seen that in terms of latency, DBR has outperformed DS and SR. DBR normally achieves a better performance rather than DS while it does not use any additional resources. Besides, SR blocks the traffic in order to avoid messages from being routed through the failed link. As we have seen, it has a great impact on increasing the message latency.

## 5. CONCLUSION

In this paper we have presented and evaluated DBR mechanism. DBR is a dynamic reconfiguration mechanism that uses a regressive deadlock recovery scheme in order to guarantee the deadlock-freedom condition in the network. Our methods can be used for any topology with various routing algorithms. DBR guarantees a deadlock-free reconfiguration based on wormhole switching (WS) and it does not require additional resources. This is the first implementation of dynamic reconfiguration method with WS. In addition, this method promises hand in messages during the reconfiguration by modifying WS to include additional flits or control signals. Evaluating results reveal that the mechanism shows substantial performance improvements over the other methods and it works efficiently in different topologies with various routing algorithms. Moreover, it can provide superior performance over SR under various traffics, especially, when the same amount of hardware is required. DBR normally achieves a better performance rather than DS while it does not use any additional resources. As for future work, we are planning to apply our approach on on-chip interconnection networks such as NoC to improve performance and decrease the power consumption.

## AUTHORS

**Majed ValadBeigi**

He received the B.S. degree from Shahid Rajaee University, Tehran, Iran, in 2008, and the M.S. degree from Shahid Beheshti University (SBU), Tehran, Iran, in 2011, both in computer engineering. He is currently working toward the Ph.D. degree from the Department of Electrical and Computer Engineering, SBU. He is a member of Scorpius simulation team which has participated in many international RoboCup Competitions. He is also an IEEE student member. His current research interests include different aspects of high-performance computing and computer-aided design with a particular emphasis on network-on-chip architectures.

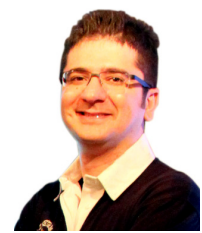

**Farshad Safaei**

He received the B.Sc., M.Sc., and Ph.D. degrees in Computer Engineering from Iran University of Science and Technology (IUST) in 1994, 1997 and 2007, respectively. He is currently an assistant professor in the Department of Electrical and Computer Engineering, Shahid Beheshti University, Tehran, Iran. His research interests are performance modelling/evaluation, Interconnection networks, computer networks, and high performance computer architecture.

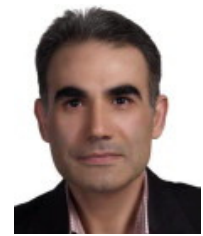

**Bahareh Pourshirazi**

She received the B.Sc. degree in Computer Hardware Engineering from Shahid Beheshti University (SBU), Tehran, Iran, in 2010. She is currently M.Sc. student in Computer Architecture Engineering in Department of Electrical and Computer Engineering, SBU, Tehran, Iran. She is a member of SBU RoboCup team which has participated in many international RoboCup Competitions. Her main Researches focus on VLSI physical design and computer-aided design with a particular emphasis on network-on-chip architectures.

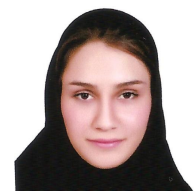